\documentstyle[preprint,prc,aps,epsf,amssymb]{revtex}

\newcommand{\bmq}{{\mbox{\boldmath $q$}}}

\begin{document}
\preprint {WIS-02/39 Sept-DPP}
\draft
\date{\today}
\title{Extraction of the static magnetic form factor and the structure 
function of  the neutron from inclusive scattering data on light nuclei}
\author{A.S. Rinat}
\address{Weizmann Institute of Science, Department of Particle Physics,
Rehovot 76100, Israel}
\maketitle
\begin{abstract}

We show that quasi-elastic inclusive electron scattering data on light nuclei
for medium $Q^2$ furnish information on $G_M^n(Q^2)$, whereas the deep-
inelastic region for large $Q^2$, provides the Structure Function 
$F_2^n(x,Q^2)$. Common to the two extractions is the possibility to 
de-convolute medium effects, which is most accurately done for light 
targets. Results are independent of the target.

\end{abstract}

\underline{Introduction.}
Most neutron observables can only indirectly be extracted from experiments 
on a nuclear medium, in which the $n$ is embedded. We discuss below the 

neutron static magnetic form factor and its Structure Function (SF).

Consider the reduced cross section for inclusive scattering of unpolarized 
electron of energy $E$ from non-oriented targets $A$ over een angle $\theta$  
\begin{eqnarray} 
  \frac{A^{-1}d^2\sigma_{eA}(E;\theta,\nu)/d\Omega\,d\nu}
  {\sigma_M(E;\theta,\nu)}= \bigg\lbrack\frac {2xM}{Q^2}
  F_2^A(x,Q^2)+  \frac{2}{M}{\rm tan}^2(\theta/2)F_1^A(x,Q^2) \bigg\rbrack
\label{a1}       
\end{eqnarray}
$F_k^A(x,Q^2)$ are two nuclear structure functions (SF), functions of 
$Q^2=\bmq^2-\nu^2$ ($\nu,\bmq$ are the energy-momentum transfer) and the 
Bjorken variable $x=Q^2/2M\nu$, with range $0\le x\le A$ ($M$ is the 
nucleon mass). Of crucial importance is a relation between the SF of nuclei 
and of nucleons. For instance (for $Z=N$) \cite{gr} 
\begin{eqnarray}
F^A_k(x,Q^2)&=&\int_x^A\frac {dz}{z^{2-k}} [f^{PN,A}(z,Q^2)
\bigg [F_k^p \bigg (\frac {x}{z},Q^2\bigg )
+F_k^n\bigg (\frac {x}{z},Q^2\bigg ) \bigg ]\bigg /2
\label{a2}
\end{eqnarray}
The two SF are related by $f^{PN,A}$, the SF of a nucleus, composed of 
point-nucleons. A standard calculation of $F_k^A$ thus requires data on 
$F_k^p$, an assumed form for $F_k^n$ and in addition, a computed, unphysical 
$f^{PN,A}$. 

We separate $F_k^N$ in NE ($\Gamma^*+N\to N$) and NI parts ($\gamma^*+N\to$ 
hadrons, partons), leading to the corresponding components $F_k^{A,NE}$  
\cite{commar} ($\eta=Q^2/4M^2$)
\begin{mathletters}
\label{a3}
\begin{eqnarray}
F_1^{A,NE}(x)&=&\frac {f^{PN,A}(x)}{4} G_d^2
[(\alpha_p \mu_p)^2+(\alpha_n \mu_n)^2]
\label{a3a}\\
F_2^{A,NE}(x)&=& \frac {xf^{PN,A}(x)G_d^2}{2(1+\eta)} \bigg [(\alpha_p 
\gamma)^2+ \bigg (\frac{\mu_n \eta}{1+5.6\eta}\bigg )^2
+\eta [(\alpha_p\mu_p)^2+(\alpha_n \mu_n)^2] \bigg ],
\label{a3b}
\end{eqnarray}
\end{mathletters}
where  reference to $Q^2$ has been dropped.
Instead of the actual static electromagnetic form factors $G^N_{M,E}(Q^2)$,
we use in Eq. (\ref{a3}) their deviations from the standard dipole form 
\cite{sill,brash,mjon}.
\begin{mathletters}
\label{a4}   
\begin{eqnarray}
\alpha_N \equiv G_M^N/\mu_NG_d &&\,\,;N=p,n
\label{a4a}\\
\gamma \equiv\frac {\mu_p G_E^p}{G_M^p}&=&\frac{G_E^p}{\alpha_p G_d}
\label{a4b}\\
\gamma=1 +\theta(Q^2-0.3)\approx&&[1-0.14(Q^2-0.3)]\,\, ; Q^2\lesssim 5.5
\label{a4c}
\end{eqnarray}
\end{mathletters}
For $G_E^n$ we use the Galster parametrization \cite{galster}. 
Nuclear NI components completely dominate cross sections on the inelastic 
side $x\lesssim 1$ of the QEP, while  for $x\gtrsim 1$ NE$>$NI. Those regions
will be treated separately.
 
\underline{Quasi-elastic region $x\lesssim 1\,$: $G_M^n$.} Consider 
first the $x,Q^2$ dependence of $F_k^{A,NE}(x,Q^2)$. The latter is primarily
due to the form 
factors in Eqs. (\ref{a3}), which decrease with growing $Q^2$. The 
$x$-dependence resides in $f^{PN,A}(x,Q^2)$, which sharply decreases with 
growing $|1-x|$ away from  the QEP at $x\approx 1$. From the above one  
concludes that ln[$\sigma^{A,NE}/A]$ grows with increasing $\nu$ 
(decreasing $x$ for fixed $Q^2$), while in general for $A\ge 12$ there 
is a mere break in the slope in the
QE region $|1-x|\ll 1$  for $A\ge 12$  (Fig. 1a) \cite{arr1}.   

The unusual structure of the lightest nuclei, causes $f^{PN,A}(x,Q^2)$ 
to be narrow and sharply peaked. With no interference of NI, the above 
change in slope may develop into a QE peak, as observed for D \cite{nicu} 
and $^4$He \cite{ne3} (Fig. 1b). For the same targets one can compute with 
great precision ground states \cite{bench} and non-diagonal target density 
matrices in the expression for $f^{PN,A}$ \cite{rtd,viv}.

Under the above circumstances one tends to ascribe the total cross sections 
on the  elastic side $x\gtrsim 1$ to NE. With $G_{E,M}^p$ known and small 
$G_E^n$, this enables the extraction of $G_M^n$ from NE. Tests for the 
above allocations are: i) Around $x\lesssim 1$, $\,\sigma^A/\sigma_M
\propto f(x,Q^2)$, i.e. of a bell shape in $1-x$. ii) $G_M^n(Q^2)$ should 
be independent of the value of the individual $x$ from which the one
extracts $G_M^n$. iii) Idem for the chosen target.

Our analysis comprises older $D$ data, where separation into transverse and 
longitudinal SF, with the former ${\cal R}_T\propto [G_M^p]{^2}+[G_M^n]{^2}$ 
\cite{lung}. Although direct and simple, it requires high-quality data in 
order to allow an accurate Rosenbluth separation and to obtain a precise 
$G_M^n$. Table I summarizes all our findings for $\alpha_n(Q^2)$ while 
Fig. 2 shows all $\alpha_n(Q^2)$, extracted thus far. Our values follow the trend of 
previously measured values and adds points for intermittent $Q^2$. Hardly 
any target dependence has been detected.

\underline{The deep-inelastic region, $\,x\ll 1$: extraction of 
$F_2^n(x,Q^2)$}. That region is dominated by NI. We focus on $F_2^n(x,Q^2)$, 
commonly estimated from the $'$primitive$'$ ansatz $F_2^n
\approx 2F_2^D-F_2^p$, which is only reliably for $x\lesssim 0.3$. Instead 
of a vehicle to compute $F_k^A$, we now consider  Eq. (\ref{a2}) in the 
inverse sense: Can one, with data on $\sigma^A$, Eq. (\ref{a1}), known 
$F_2^p$ and computed $f^{PN,A}$ extract $F_2^n(x,Q^2)$? 

Virtually all previous methods addressed a D target (e.g. \cite{meln}). We 
outline and apply a method \cite{rtf2n}, which with sufficient kinematics 
available \cite{arr1,nicu}, is applicable to all targets.(see Refs. 
\onlinecite{afn,pace} for treatments of isobar pairs). Again a test is an 
outcome, independent of $A$. As to $F_2^A$, in order to separate it from
$F_1^A$, one needs in addition to cross sections, an assumption on 
$R^{-1}(x,Q^2)+1\propto 2xF_1^A(x,Q^2)/F_2^A(x,Q^2)$. Alternatively, one 
may for every data point determine a relative deviation of theory and data, 
and ascribe it in equal measure to the two SF. The procedure produces 
quasi-data for $F_2^{A;qd}$.
 
All modern data thus far \cite{arr1,nicu} appear to yield $F_2^A$ in disjoint 
$x,Q^2$ regions, whereas the inversion of Eq. (\ref{a2}) requires data 
over a large $x$-range for the same $Q^2$. Even with careful binning and/or 
interpolation, we could only construct a single set for $Q^2\approx$ 
3.5 GeV$^2\,,x\gtrsim 0.55$, which $x$-range misses a crucial part of the DI 
region. Fortunately, one can use the fact, that, independent on $Q^2$, 
$F_2^p(x,Q^2)\approx 0.32$ for $x\approx 0.16$. Eq. (\ref{a2}) then proves 
the same for $F_2^A(x,Q^2)$, permitting extrapolation into the vital DI region.   

We have used several inversion methods, all based on a parametrization 
\begin{eqnarray}
F_2^n(x,Q^2)=F_2^n(x,Q^2;d_k)&=&C(x,Q^2;d_k)F_2^p(x,Q^2)
\nonumber\\
C(x,Q^2;d_k)&=&\sum_{k\ge 0} d_k(Q^2)(1-x)^k,
\label{a5}
\end{eqnarray}
with  mildly constrained parameters. First we take $C(0)=1$, ensuring a 
finite outcome for the Gottfried sumrule $S_G(Q^2)=\int_0^1\frac {dx}{x}
[F_2^p(x,Q^2)-F_2^n(x,Q^2)]$. Next we exploit the above $'$primitive$'$ 
ansatz for, say, $x=0.2$. For the simplest choice $k_{max}=2$ only one 
parameter is left, e.g. $d_0=C(1)$. It moreover proved useful to parametrize 
$F_2^p$ as follows
\begin{mathletters}
\begin{eqnarray}
\label{a6}
F_2^p(x,Q^2)&=x^{-a^2}&\sum_{m\ge 1} c_m(1-x)^m ;x\ge 0.02
\label{a6a}\\
&=&0.42~~~~~~~~~~~~~~~~;x\le 0.02
\label{a6b}
\end{eqnarray}
\end{mathletters}
In the region $0.02 \lesssim x\lesssim 0.9$, the above practically coincides 
with the standard parametrization \onlinecite{amad}. Fig. 3 shows our results for 
$C, F_2^n$ for fixed  $Q^2=3.5$ GeV$^2$ and given $F_2^p$. The band in $C$ 
reflects results from several inversion methods and from different targets 
D,C,Fe. The value of $C$ at the elastic point $x=1$  has been the subject of 
several estimates with results, marked by small horizontal lines. All those, as 
well as our $C$, assumed smooth, i.e. resonance-averaged behaviour of $F_2^N$ 
(cf. lower part of Fig. 3). 

The above is an undesired feature of averaging: the lowest inelastic 
threshold of $F_2^N(x,Q^2)$, occurs at a mass $M+m_{\pi}$, or equivalently, 
at $x_{thr}(Q^2)=[1+2Mm_{\pi}/Q^2]^{-1}$. In particular $x_{thr}(3.5)\approx 
0.93$, which is marked in Fig. 3 by a vertical line. For  $x_{th}<x<1\,\,,
F^N(x,Q^2)$ is strictly 0.  In particular the mention prediction of $C$ out 
to the elastic border, merely reflects the different approach to 0 of
the $p,n$ SF. As a consequence $C(x \to 1)$ is due to purely NE parts of 
$F_2^N$, and equals (cf. Eq. (\ref{a3b}))
\begin{eqnarray}
{\lim_{x \to 1}C(x,Q^2)}=\bigg [
\frac{\mu_n\alpha_n(Q^2)}{\mu_p\alpha_p(Q^2)} \bigg ] \bigg [1+
\frac{4M^2}{Q^2}\bigg (\frac {\gamma(Q^2)}{\mu_p}\bigg )^2 \bigg]^{-1},
\label{a7}
\end{eqnarray}
From Eqs. (\ref{a4}), (\ref{a7}) one then $computes$
\begin{eqnarray}
C(x=1,3.5)\approx 0.61, 
\label{a8} 
\end{eqnarray}
surprisingly close to the $extracted$ value as the ratio of the two $F_2^N$,
which  tend to 0 in a different way for  $x\to 1$. More extensive reports 
can be found in Refs. \cite{rtg,rtf2n}.

\underline{Acknowledgements}; Part of this work has been done in collaboration with
M.F. Taragin and M. Viviani.

{Figure captions}
 
Fig. 1a,b. Partial data and predictions for inclusive cross sections 
($E=4.045$ GeV, $\theta=15^{\circ},23^{\circ},30^{\circ}$) on D,Fe.

Fig. 2. $\alpha_n=G_M^n/\mu_n G_d$ as function of $Q^2$. Shown are some
previous representative results. Filled squares, diamonds, triangles and 
stars are our results.   

Fig. 3.  The ratio $C(x,3.5)=F_2^n(x,3.5)/F_2^p(x,3.5)$ for
$Q=3.5\,$ GeV$^2$  from data on D, C, Fe. The drawn line corresponds to
$C(1)=0.54$ and the band represents the spread from averages over different 
targets and methods. The numbers on the right abscissa are standard quark 
model and QCD predictions for $C(1)$ with 0.61, the $NE$ limit (\ref{a7}).

\begin{table}
\caption {Extraction of $\alpha_n(Q^2)$ from QE inclusive scattering data on 
D, $^4$He. Columns give target, beam energy $E$, scattering angle $\theta$,
ranges of Bjorken $x$ and $Q^2$, range of SF of target composed of
point-nucleons and (between brackets) its maximal value. The last column
gives $\alpha_n(Q^2)$ with deviations from average over the considered
$x$-intervals.}

\begin{tabular} {c|c|c|c|c|c|c|}
\hline
%{c| c@{\qquad}c| c@{\qquad}cl c@{\qquad}cl c@{\qquad}cl
%c@{\qquad}cl c@{\qquad}cl c@{\qquad}cl} 
% &  \multicolumn{1}{c}{target}
% &  \multicolumn{1}{c}{E}
% &  \multicolumn{2}{c}{$\theta,Q^2}
% &  \multicolumn{2}{c}{$x,x f^{PN,A}(x,Q^2)$} 
% &  \multicolumn{1}{c}{$Q^2$}
% &  \multicolumn{2}{c}{$F_2^A(x,Q^2)$}
% &  \multicolumn{2}{c}{$\alpha_n(Q^2$}\\ 

target    & $E$ (in GeV) & $\theta$  & $x$ & $Q^2({\rm in\,GeV}^2)$ & 
$f^{PN,A}(x,Q^2)$ & $\alpha_n(Q^2)$\\  
\hline
 $^4$He$^{\cite{ne3}}$ & 2.02  & $20^{\circ}$ & 1.125-0.848 & 0.444-0.430 
& 0.97-1.49 (1.49)  & 0.988$\pm 0.055$ \\ 
 -      & 3.595 & $16^{\circ}$ & 1.125-0.930 & 0.887-0.864
&  1.16-1.90 (1.90) & 0.967$\pm 0.028$ \\            
 -      & 3.595 & $20^{\circ}$ & 1.095-0.925 & 1.295-1.250
&  1.44-2.16 (2.16) & 0.988$\pm0.018$ \\ 
\hline
   $D^{\cite{nicu}}$  & 4.045 & $15^{\circ}$ & 1.131-0.953 & 0.988-0.972
&  1.31-3.65 (4.30)  & 1.039 $\pm0.020$  \\   
   -    & 4.045 & $23^{\circ}$ & 1.079-0.978 & 1.976-1.929
&  2.44-5.18 (5.18)   & 1.062 $\pm0.009$ \\
\hline  
   $D^{\cite{lung}}$  & 5.507 &  $15.2^{\circ}$ & 1.063-0.978 & 1.769-1.741
&  2.89-5.04 (5.31) & 1.047 $\pm0.019$  \\
   -    & 2.407 & $41.1^{\circ}$ & 1.081-0.957 & 1.803-1.721
&  2.37-4.89 ((5.32)  & 1.048 $\pm0.007$  \\                        
   -    & 1.511 & $90.0^{\circ}$ & 1.059-0.977 & 1.812-1.728 
 &  3.21-4.79 (5.26)  & 1.057 $\pm0.009$  \\                        
  ${{\cal R}_T^{D,NE}}^{\cite{lung}}$ &3.809 & $20^{\circ}$ &1.141-0.962 &
$<Q^2>$=1.75 
&  1.79-3.38 (5.31) & 1.004$\pm0.014 \,\bigg (1.052^{\cite{lung}}\bigg )$ \\
\hline   
  $D^{\cite{lung}}$  & 5.507 & $19.0^{\circ}$ & 1.104-1.000 & 2.561-2.501
&  1.69-5.65 (5.98)  & 1.030 $\pm0.016$  \\                        
     -& 2.837 & $45.0^{\circ}$ & 1.101-0.991 & 2.613-2.500
&  1.69-5.91 (5.94)  & 1.031 $\pm0.018$  \\                        
   -  & 1.968 & $90.0^{\circ}$ & 1.064-0.984 & 2.608-2.474 
&  3.06-5.71 (5.90)  & 1.078 $\pm0.027$  \\                         
  ${{\cal R}_T^{D,NE}}^{\cite{lung}}$ &5.016 & $20^{\circ}$ & 1.068-0.940& 
$<Q^2>$=2.50 
&  2.92-4.16 (5.94)  & 0.986 $\pm0.014 \,\bigg (1.014^{\cite{lung}}\bigg )$ \\
\hline
  ${{\cal R}_T^{D,NE}}^{\cite{lung}}$ &5.016 & $20^{\circ}$ & 1.051-0.958& 
$<Q^2>$=3.25                 
&  3.50-6.15 (6.43)  &0.940$\pm0.013 \, \bigg (0.967^{\cite{lung}}\bigg )$ \\
\hline 
  ${{\cal R}_T^{D,NE}}^{\cite{lung}}$ &5.016 & $20^{\circ}$ & 1.079-1.038& 
$<Q^2>$=4.00 
&  3.80-6.20 (6.50)  &0.830$\pm0.016 \, \bigg (0.923^{\cite{lung}}\bigg )$ \\ 

\end{tabular}
%\hline

%\label{Table I}
\end{table}

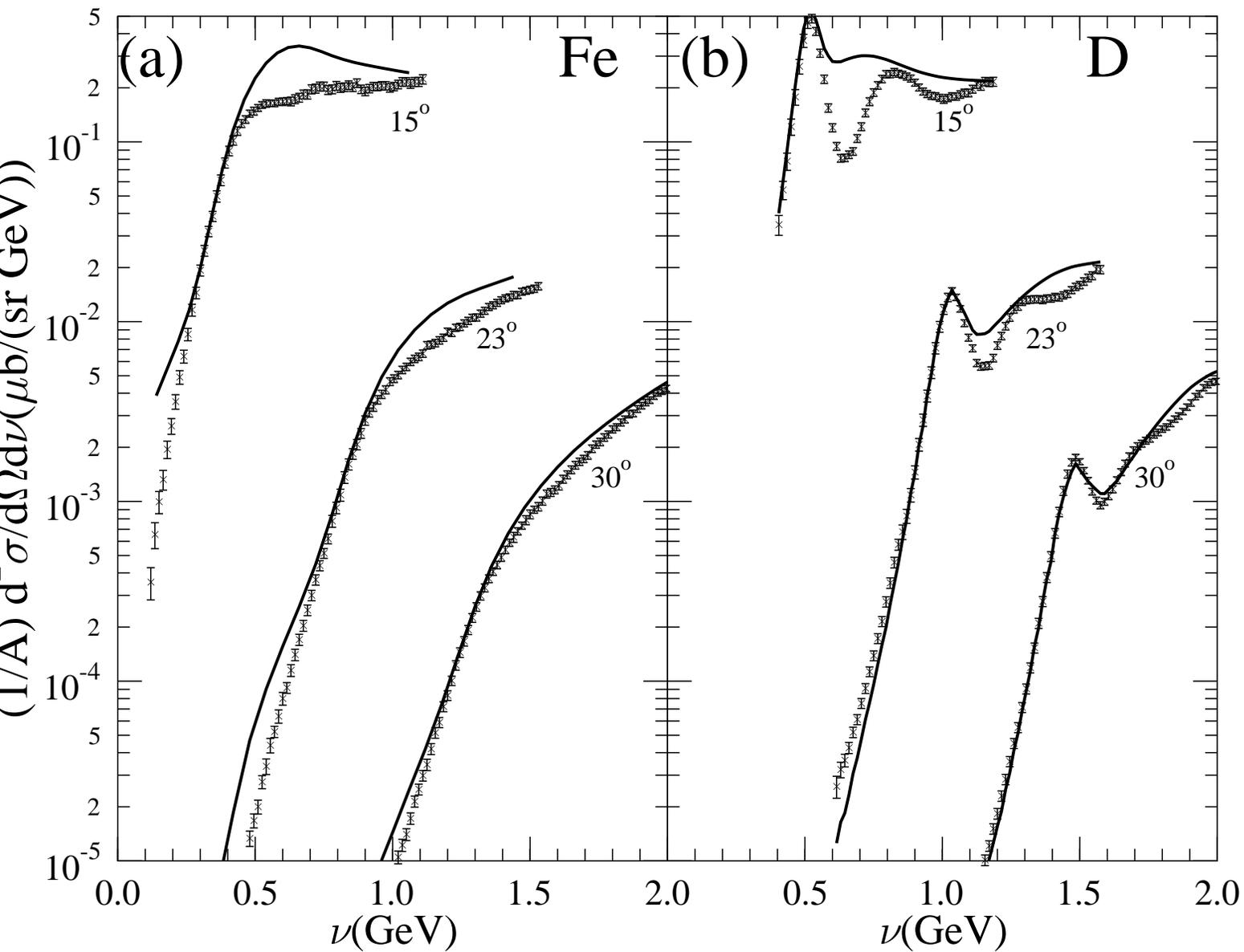
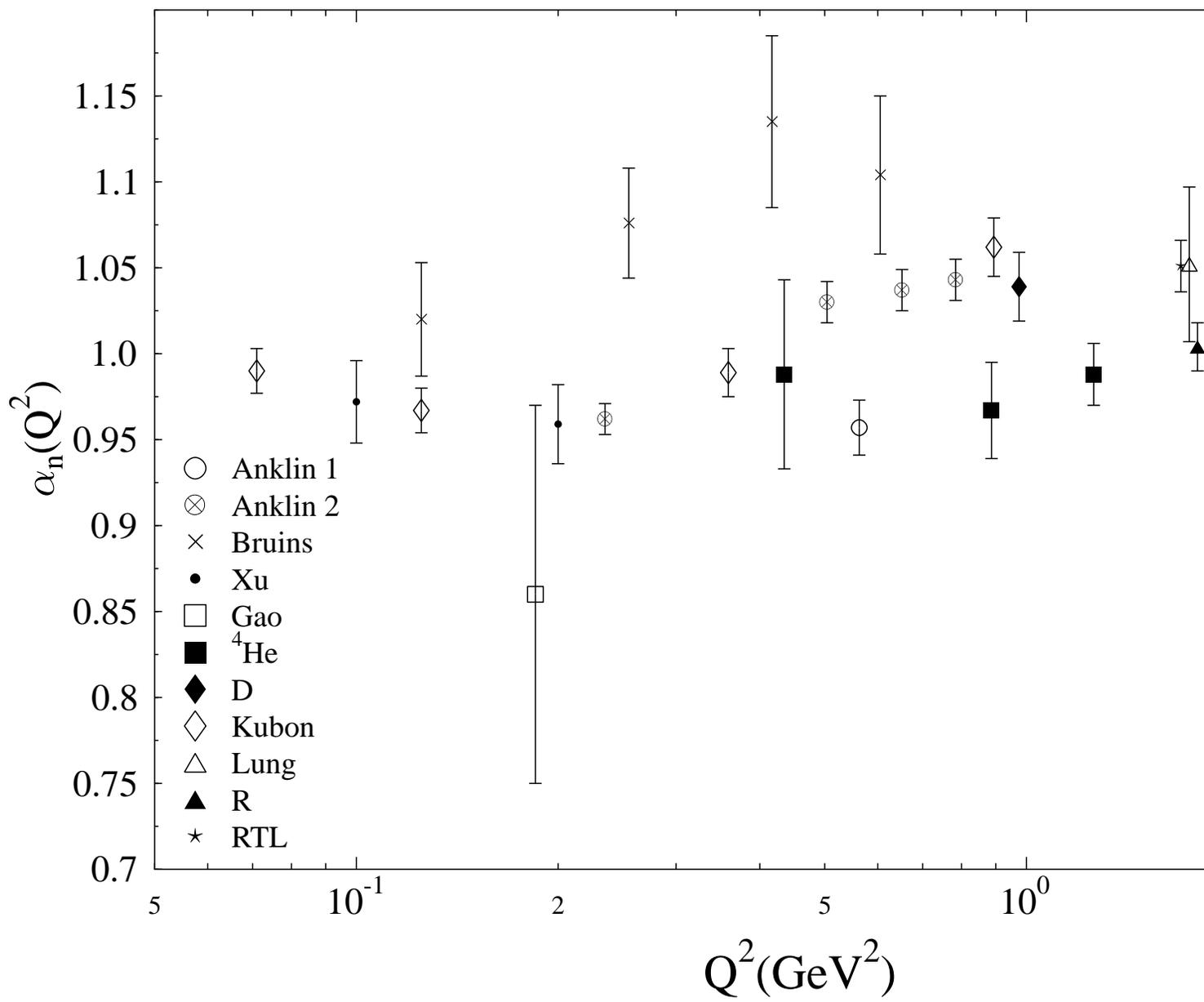
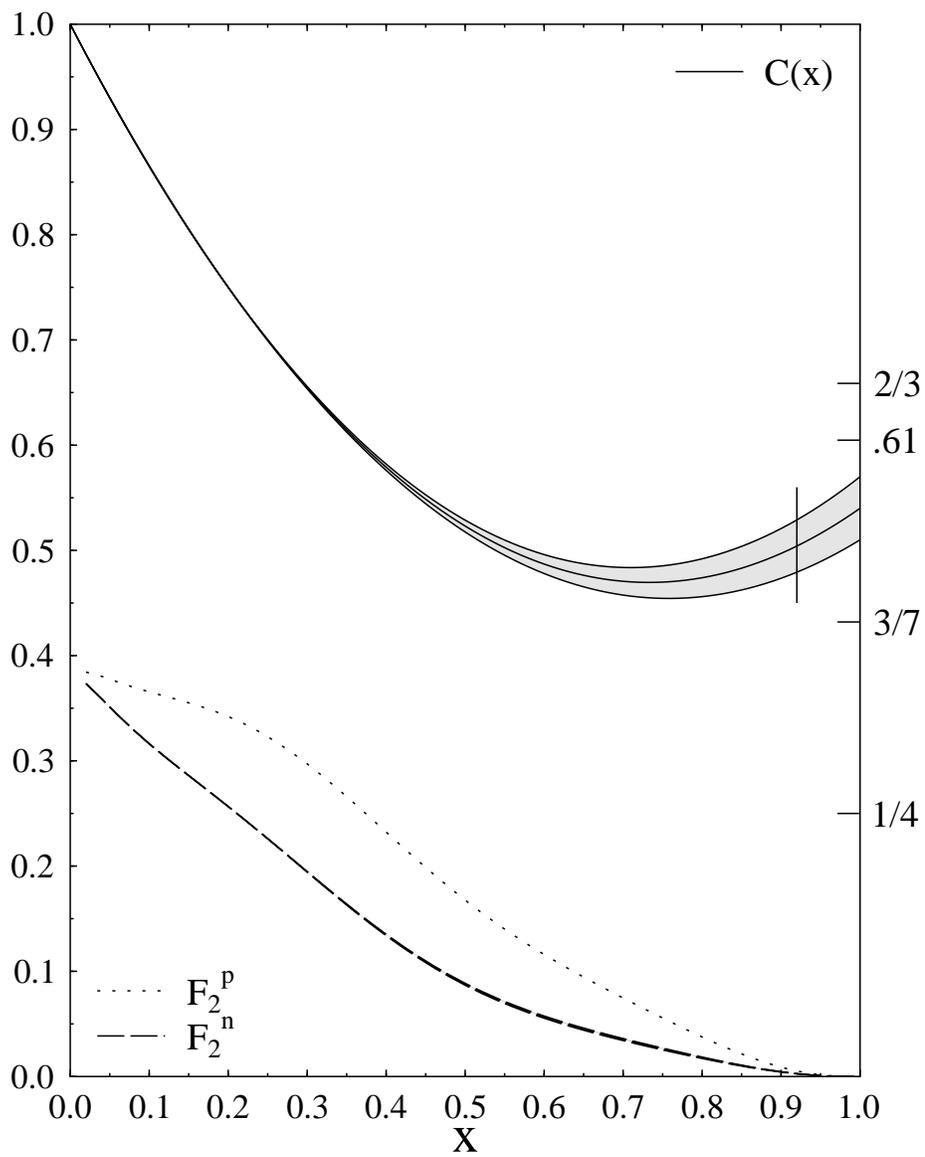
\end{document}